# Higgs Measurement at $e^+e^-$ Circular Colliders

Manqi RUAN[a, b]

[a] *CERN, 40-b1-06, Geneva, Switzerland*
[b] *Institute of High Energy Physics, 19 B Yuquan Road, Beijing, China*

**Abstract**

Now that the mass of the Higgs boson is known, circular electron positron colliders, able to measure the properties of these particles with high accuracy, are receiving considerable attention. Design studies have been launched (i) at CERN with the Future Circular Colliders (FCC), of which an $e^+e^-$ collider is a potential first step (FCC-ee, formerly caller TLEP) and (ii) in China with the Circular Electron Positron Collider (CEPC). Hosted in a tunnel of at least 50 km (CEPC) or 80-100 km (FCC), both projects can deliver very high luminosity from the Z peak to HZ threshold (CEPC) and even to the top pair threshold and above (FCC-ee).

At the ZH production optimum, around 240 GeV, the FCC-ee (CEPC) will be able to deliver 10 (5) ab$^{-1}$ integrated luminosity in 5 (10) years with 4 (2) interaction points: hence to produce millions of Higgs bosons through the Higgsstrahlung process and vector boson fusion processes. This sample opens the possibility of sub-per-cent precision absolute measurements of the Higgs boson couplings to fermions and to gauge-bosons, and of the Higgs boson width. These precision measurements are potentially sensitive to multi-TeV range new physics interacting with the scalar sector. The ZH production mechanism also gives access to the invisible or exotic branching ratios down to the per mil level, and with a more limited precision to the triple Higgs coupling. For the FCC-ee, the luminosity expected at the top pair production threshold ($\sqrt{s} \sim 340\text{-}350$ GeV) further improves some of these accuracies significantly, and is sensitive to the Higgs boson coupling to the top quark.

*Keywords*: Higgs factory, Circular Collider, FCC-ee, CEPC

## 1. Introduction

The Higgs boson is not only the last found piece of the Standard Model (SM) [1, 2], but also an extremely strange object. It is the only scalar particle in the SM, and it is also responsible for most of the SM theoretical difficulties and defects. The Higgs boson might be a portal leads to more profound physics models and even physics principles. Therefore, a Higgs factory that can precisely determine the properties of the Higgs boson is an important future step in the high-energy physics exploration.

The LHC itself is already a Higgs factory. It not only discovered the Higgs boson, but also demonstrated that the discovered Higgs boson is highly likely to be the one predicted by the SM. The future HL-LHC program will, with 2 orders of magnitude higher integrated luminosity, highly enhance the understanding of this mysterious boson.

Higgs measurement at a proton collider is limited by the Higgs width measurements. The natural width of a 125 GeV Higgs boson is only 4.2 MeV, while the best precision achieved with current LHC data is of o(10) MeV [3]. The Higgs width measurement is an essential ingredient to determine the partial width and then the coupling constants. On the contrary, the electron-positron collider has a remarkable capability of precision measurement on absolute Higgs couplings and exotic decay branching ratios.

At electron-positron collider, the Higgs bosons are produced through Higgsstrahlung and vector boson fusion processes (Figs. 1 and 2). In the Higgsstrahlung process, the leading production process at center-of-mass energy of 240 - 250 GeV, the Higgs boson is produced in association with a Z boson. Therefore, the Higgs signal can be tagged with the Z boson decay (from the mass of its decay products, and from the recoil mass spectrum to the Z boson) – especially if the Z boson decays into a pair of leptons that can be clearly tagged and precisely measured. In this way, the Higgs signal is tagged model independently as no information from the Higgs decay final states is used. The inclusive production cross section of the Higgs boson is then measured through the total number of Higgsstrahlung events, from which the absolute coupling g(HZZ) can be inferred. The measured g(HZZ) then serves as the anchor for the Higgs width and the other coupling constants measurement.

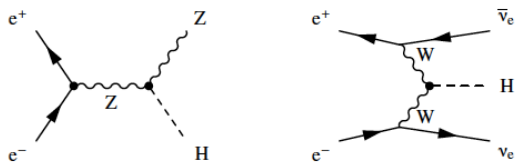

Fig. 1. Leading Higgs production Feynman diagrams at electron positron collider with c.o.m energy of 240-250 GeV. Left: Higgsstrahlung process; Right: W fusion process

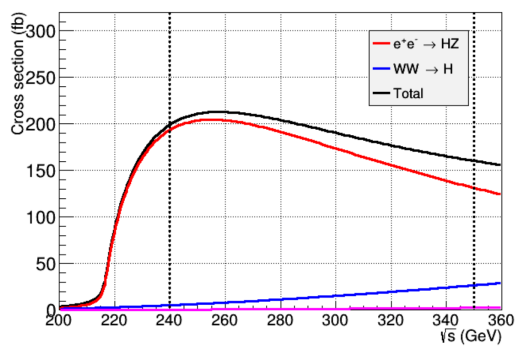

Fig. 2. Differential cross section of Higgs generation process at electron positron collider

Once determined, the Higgs couplings can directly be compared with the Lagrangian: of the SM or the new physics. A large variety of new physics models predict deviations of the coupling constant from the SM values, and the amplitude of the deviation is usually proportional to $\Lambda^{-2}$, where $\Lambda$ represent the energy scale of new physics. In general, deviations at per-cent level would give strong indications on the new physics models in the TeV range. The better accuracy, the higher the testable energy scale. Accuracy of a per-cent or preferably below has become the physics objective of the next generation of electron positron Higgs factory [4].

The Higgs measurement program at the two proposed circular electron-positron collider facilities, FCC-ee [5] and CEPC [6], is now given. In Section 2, the collider layout for these two facilities, and the collider parameters, are briefly presented. In Section 3 the Higgs programs and the expected accuracies at these facility are summarized. A discussion on the constrains of g(Htt) and g(HZZ) are presented in Section 4. The complementarity of a proton-proton collider and electron positron collider is exemplified in Section 5.

## 2. Collider parameters: centre-of-mass energy, Luminosity and Yields

There are two basic types of the electron-positron colliders: linear or circular. Nowadays, the former includes the International Linear Collider (ILC) [7] and the Compact Linear Collider (CLIC) [8] while the latter is represented by the FCC-ee and CEPC (Fig. 3).

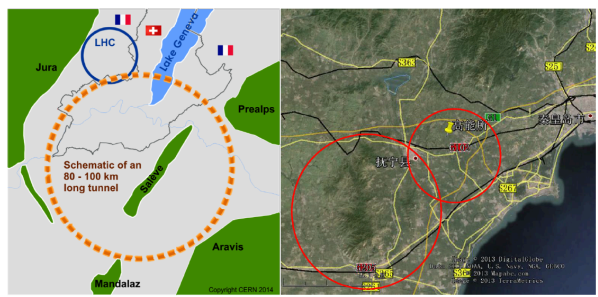

Fig. 3. Example layouts of FCC-ee project (Left) and CEPC project (Right). The FCC-ee will be located at the CERN site while the location of the CEPC is not yet fixed. An example layout near the city of Qinhuangdao is demonstrated in the right plot. The two circles in the right plot have 50 km and 100 circumferences.

The FCC-ee and CEPC will be operated at several centre-of-mass energies, of which √s = 240-250 GeV is suitable for the Higgs boson precision measurements. Dedicated runs at the Z pole and the WW production threshold are equally important, as they provide excellent electroweak precision measurements. Data at the Z pole also provide detector calibration to reduce the detector systematic uncertainties to a negligible level for the Higgs measurements. The FCC-ee design and programme also include a very important run at the top production threshold, towards a precise measurement of the top quark mass, essential for the interpretation of the electroweak measurements. The physics at the Z pole, and at the WW and the top threshold are described in another presentation at this conference [9].

Though limited by the synchrotron radiation power, circular colliders are able to deliver high to very-high luminosity through a large collision rate and multiple interaction points. For the Higgs run at 240-250 GeV, both facilities have expected luminosity of the order of $10^{34}$cm$^{-2}$s$^{-1}$: The FCC-ee reaches $5*10^{34}$cm$^{-2}$s$^{-1}$ per interaction point, while CEPC reaches $2*10^{34}$cm$^{-2}$s$^{-1}$ per interaction point. The difference is partly originated to the collider circumference, and the fact that FCC-ee uses double storage ring while CEPC use single storage ring in their base line design (The cost control has being given a high priority in the CEPC project). The instance luminosities of these facilities are compared to other options (in Fig.3).

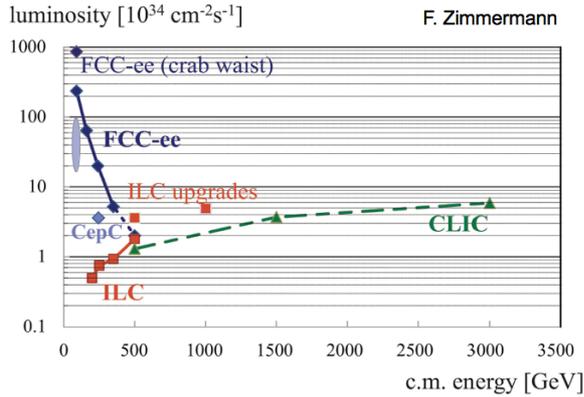

Fig. 4. Instance luminosity for different electron-positron facilities.

In current baseline, FCC-ee proposed to run 5 years at 240-250 GeV and then 5 years at the top threshold (340-350 GeV), while CEPC will stay at 240-250 GeV for 10 years. Giving 4/2 interaction points, FCC-ee/CEPC will be able to produce 2M/1M Higgs bosons with 5/10 years operation (at CEPC 1 year operation is expected to be $1.25*10^7$ second). The FCC-ee is also plan to accomplish its Higgs measurement through the top runs, where the precision on the Higgs total width will be reduced by a factor three, down to better than a per-cent, which in turn further constrains the Higgs couplings.

## 3. Main observables and expected accuracy

At electron positron collider, the Higgs can be measured by three methods. The first method is through the recoil mass to the associated Z boson: if the Z boson decays into visible final states, especially a pair of electrons or muons. The recoil mass spectrum give us the access to the Higgs mass and inclusive cross section σ(ZH) thus g(HZZ). Explained in the introduction, measurement through the recoil mass spectrum is unique and provides model independence of all subsequent measurements.

The second method is through the tagging of Higgs decay final states. This method allows the number of Higgs events to be counted in various final states, thus measures the product of Higgs production cross section and the Higgs decay branching ratio to a given final states: σ(ZH)×Br(H → X), σ(vvH)×Br(H → X). σ(vvH) refers to the cross section of W-fusion process, while X denote the possible final states: either SM modes (as WW, ZZ, bb, cc, gg, $\tau\tau$, $\gamma\gamma$ etc) or Beyond SM modes (for example invisible decay).

The third method is through the differential distributions. This method determines the quantum numbers of Higgs bosons such as the Spin, Parity and CP.

The measurements through the first and second methods are simple counting measurements. Giving the expected luminosity at FCC-ee and CEPC, the direct observables and their expected accuracy at 240-250 GeV Higgs run is presented in Table 1 [10].

Table. 1.
List of the main observables and expected accuracy at FCC-ee and CEPC with 2 Million/1 Million Higgs boson respectively

|  | FCC-ee 240GeV | CEPC 250GeV |
|---|---|---|
| Higgs mass | - | 5.4 MeV |
| σ(ZH) | 0.4% | 0.7% |
| σ(ZH)×Br(H → bb) | 0.2% | 0.4% |
| σ(ZH)×Br(H → cc) | 1.2% | 2.1% |
| σ(ZH)×Br(H → gg) | 1.4% | 1.8% |
| σ(ZH)×Br(H → WW) | 0.9% | 1.3% |
| σ(ZH)×Br(H → ZZ) | 3.1% | 5.1% |
| σ(ZH)×Br(H → ττ) | 0.7% | 1.2% |
| σ(ZH)×Br(H → γγ) | 3.0% | 8.0% |
| σ(ZH)×Br(H → μμ) | 13% | 18% |
| σ(vvH)×Br(H → bb) | 2.2% | 3.8% |

These expected accuracies are acquired with different approaches. The FCC-ee study has so far conservatively used a full simulation of the CMS detector, likely to be not the optimal configuration in an electron-positron collision environment. The CEPC results are based on the ILD detector [11], one of the baseline designs for the linear electron-positron colliders (ILC/CLIC). These CEPC numbers are extrapolated from ILC results and confirmed/updated with independent Fast/Full analyses.

Though measured with very different detector and follow different analysis approach, the accuracy in Table 1 basically statistically agrees with each other (FCCee has twice more statistic). The accuracy of σ(ZH)×Br(H → γγ) at FCCee is much better because the intrinsic photon energy resolution at CMS detector is roughly a factor of two better than that of ILD at the corresponding energy range.

The top run of FCC-ee brings not only more Higgs bosons but also a much larger sample of the W fusion event. With which, a much better accuracy on the σ(ννH)×Br(H → bb) can be achieved: improved from 2.2% to 0.6%. This measurement is one of the main inputs to determine the Higgs width [3, 10], therefore, the width measurement at FCC-ee (1.0%) is significantly better than CEPC (3.5%).

Since the electron positron collider determines both σ(ZH)×Br(H → X) and σ(ZH), the absolute Higgs decay branching ratios are measured. In addition, the Higgs total width is also measured, therefore the absolute couplings between Higgs boson and its decay final states are determined.

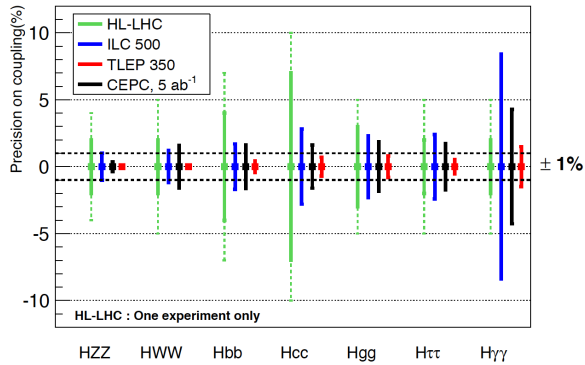

Fig. 5. Expected Higgs coupling measurement at different Higgs factories.

Fig. 5 shows the accuracy of Higgs coupling measurement at different facilities:

The high luminosity LHC: HL-LHC, 3 ab$^{-1}$ integration luminosity at 14 TeV centre of mass proton collision;

ILC 500: electron positron collision, with 250 fb$^{-1}$ at 250 GeV center of mass energy, and 500 fb$^{-1}$ at 500 GeV;

TLEP 350: electron positron collision with 10 ab$^{-1}$ at 240 GeV and 2.6 ab$^{-1}$ at 350 GeV;

CEPC: electron positron collision with 5 ab$^{-1}$ at 250 GeV.

In fact, Fig. 5 is an update of Fig. 12 in Ref. [7] with the CEPC results.

As mentioned above, assumptions are needed to extract the Higgs couplings from the measurements made at a proton collider. Two assumptions are used to get the HL-LHC results in Fig. 5:

No Higgs exotic decay;

Derivation of the charm and top couplings are correlated (It is almost impossible to measure g(Hcc) at LHC).

The solid and dashed lines are corresponding to different scenario on the systematic control: dashed line corresponding to the same systematic errors as today, while solid line corresponding to an optimistic scenario where systematic uncertainty scales with $1/\sqrt{L}$ as and theoretical error halved.

The accuracy of Higgs measurement at electron positron collider is dominated by the statistic of the collected data. Therefore larger statistic is always appreciated. Besides the statistic effects, FCC-ee has much better g(HWW) and g(Hbb) accuracy since the Higgs width is better determined. The g(Hγγ) coupling at FCC-ee is also significantly better than CEPC because of different intrinsic ECAL resolution at different detector concepts.

## 4. Indirect measurement of g(HHH) and g(Htt)

Besides the absolute value of the couplings between Higgs boson and its decay final states, it is also of strong physics interest to determine the other two couplings: Higgs to top couplings and Higgs self couplings. The former is important to understand the naturalness problem while the latter is the cornerstone of the EWSB mechanism.

At 240-250 GeV and 340-350 GeV centre-of-mass energies, none of these two observables can be measured directly. However, the ultra high precision of electron positron colliders does provide possibilities to constrain/measure these two parameters indirectly. The effective couplings between Higgs and massless boson, gluon and photon, give constrains on the g(Htt) couplings. In addition, Figure 6 [12] shows a NLO calculation of σ(t$\bar{t}$) near the $t\bar{t}$ threshold. At this energy, $t\bar{t}$ are generated through the intermediate particles, while different assumptions leads to different behaviours. Therefore, a precise cross section measurement, or one step further, a threshold scan with enough accuracy will be able to indicate the value of g(Htt).

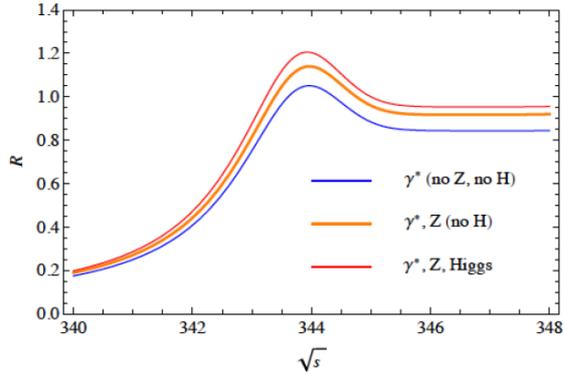

Fig. 6. Differential cross section of ttbar event near threshold

In addition, g(HHH) can be constrained from the precise measurement of g(HZZ) [13], as shown in Fig. 7. The key idea is that Higgs triplet coupling will contribute to the radiative correction to g(HZZ), which, in turn constrains the value of g(HHH).

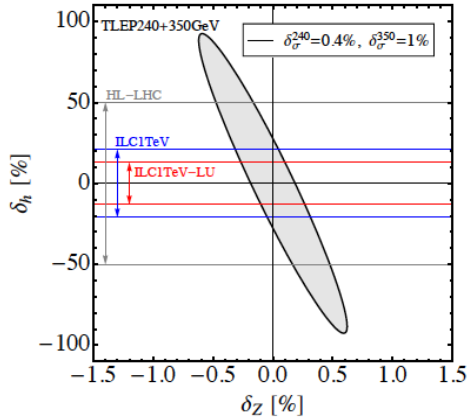

Fig. 7. Indirect constrains of g(HHH) coupling at FCC-ee

## 5. Complementary to other projects

For the Higgs measurement, the proton collider and electron positron collider plays highly complementary roles. The total production cross section at proton collider is much larger than that at electron-positron collider (2 – 4 orders of magnitude), providing the possibility to study not only the Higgs rare decays but also Higgs rare production. The latter can be more sensitive to new physics. On the other hand, the absolute measurement of Higgs couplings and width at electron-positron collider provides essential anchor to better understand the proton collision data.

Explicitly, if the Higgs boson is generated though a coupling g(HAA) and decayed through coupling g(HBB), the total number of such event is proportional to $g(HAA)^2 g(HBB)^2/\Gamma$. At proton collider where the Higgs signal is usually tagged through final states, $g(HAA)^2 g(HBB)^2/\Gamma$ can be regarded as the actual observable. Now, since the width and presumably one of the two couplings are determined at electron positron collider, the other couplings can also be inferred.

In addition, proton colliders, for example, the HL-LHC also provides direct measurement to g(HHH) and g(Htt) - with the input of electron positron collider. Using $pp \rightarrow HH \rightarrow bb\gamma\gamma$ events, the g(HHH) can be determined to 50% relative accuracy (with 1 detector) [14] while g(Htt) can even be measured to per-cent level. The $g(Htt)$ determination at proton collider is actually very interesting. For the ttH events, main signal for g(Htt) measurement, the ttZ events servers as perfect control sample – and the ratio between them, σ(ttH)/σ(ttZ), can be determined to a precision of 0.75% at HL-LHC [15].

Besides the HL-LHC, the tunnel of both CEPC and FCC-ee can be used to host a high-energy proton collider, called SPPC and FCC-hh respectively. The energy reach will be at 70TeV - 100 TeV range with an integrated luminosity of the order of 10 ab$^{-1}$. SPPC/FCC-hh has much stronger access to new physics at high-energy range, for instance, they are able to push the index of naturalness problem from per-cent level (as now at LHC) to 10$^{-4}$ level. In terms of Higgs physics, those colliders will be able to measure g(HHH) to a precision of 8% level [14]. With o(10$^{9-10}$) Higgs bosons and o(10$^{12}$) top quarks expected at such machines, there is a huge physics potential for Higgs rare decay/production process measurements and other SM measurements.

## 6. Summary

After the Higgs discovery, the precise measurements of Higgs properties become a natural and important step for the future exploration of particle physics. A circular electron positron collider that can be upgraded to proton collider with much high centre-of-mass energy is a very attractive option. Therefore, the FCC-ee and CEPC projects, circular colliders with total circumference of at least 50 km or 80 – 100 km, are proposed.

The electron positron colliders have a unique capabilities of absolute measurement of the Higgs, including the determination of Higgs total production cross section, decay branching ratios, Higgs width and ultimately the Higgs couplings. On the contrary, though proton colliders can provide lots of essential information about the Higgs boson, the precise coupling constant measurements need to refer to some theoretical assumptions, or input from measurement at other facilities.

The accuracy of Higgs measurement at electron-positron collider is completely dominated by statistic. Z pole samples at those facilities will control the detector systematic to a negligible level. Using multiple interaction points and high collision rate, circular colliders are more productive than the linear colliders as the Higgs factories. Both FCC-ee and CEPC can deliver millions of Higgs bosons, leading to per-cent, and even sub-per-cent level precision on Higgs coupling measurements. The FCC-ee has better performance than CEPC, not only because the higher luminosity, but also because FCC-ee has top threshold runs, with which Higgs width and couplings are measured to better precision. On the other hand, though giving high priority on the cost control, CEPC also reaches the key objective, per-cent level accuracy for most of the coupling constant measurements.

The FCC-ee and CEPC cannot measure directly g(Htt) and g(HHH), while these two couplings are very important for a complete understanding of the nature of Higgs field and new physics at higher energy scale. However, even only use low energy runs, electron-positron colliders can indirectly constrain/determine these two couplings. On the other hand, proton colliders (HL-LHC, SPPC, FCC-hh) plays a very different, yet highly complementary role to the electron positron collider, using the input from electron positron collider, proton colliders can directly measure g(Htt) and g(HHH), meanwhile, provide huge potential in measuring Higgs rare decay as well Higgs rare production.

We hope the studies introduced in this manuscript have already demonstrated the physics potential at these circular colliders. On the other hand, dedicated reconstruction, analysis studies and detector optimization progresses are expected to further improve these results.

**Acknowledgments**

I would like to thank P. Janot, A. Blondel and M. Klute for their kind invitation to this talk and the initiative discussions with them.

**References**

[1] ATLAS Collaboration, G. Aad et al., Observation of a new particle in the search for the Standard Model Higgs boson with the ATLAS detector at the LHC, Phys. Lett. B 716 (2012) 1–29, arXiv:1207.7214 [hep-ex].
[2] CMS Collaboration, S. Chatrchyan et al., Observation of a new boson at a mass of 125 GeV with the CMS experiment at the LHC, Phys. Lett. B716 (2012) 30–61, arXiv:1207.7235 [hep-ex].
[3] F. Caolo, K. Melnikov, Constraining the Higgs boson width with ZZ production at the LHC, arXiv:1307.4935 [hep-ph]
[4] M. Peskin, Comparision of LHC and ILC Capabilities for Higgs Boson Coupling Measurements, arXiv:1207.2516 [hep-ex]
[5] http://cern.ch/fcc-ee
[6] http://cepc.ihep.ac.cn
[7] ILC Technical Design Report http://www.linearcollider.org/ILC/Publications/Technical-Design-Report
[8] CLIC Conceptual Design Report http://project-clic-cdr.web.cern.ch
[9] R. Tenchini, Precision Electroweak Measurements at FCC-ee, presentation at ICHEP 2014, Valencia, Spain
[10] M. Bicer et al., First Look at the Physics case of TLEP, arXiv:1308.6176 [hep-ex]
[11] ILD Detailed Baseline Design http://www.linearcollider.org/ILC/physics-detectors/Detectors/Detailed-Baseline-Design
[12] M. Beneke, Top pair production near threshold, presentation at 7[th] FCC-ee physics workshop, June 2014, CERN. See also arXiv:1312.4791[hep-ph]
[13] M. McCullough, An indirect model-dependent probe of the Higgs Self-Coupling, arXiv:1312.3322 [hep-ph]
[14] S. Dawson, A. Gritsan et al., Higgs Working Group Report of the Snowmass 2013 Community Planning Study, arXiv: 1310.8361 [hep-ex]
[15] M. L. Mangano, ttH/ttZ as a precision probe of the top Yukawa coupling, presentation at the 1[st] Future Hardon Collider Workshop, May 2014, CERN